\documentclass{aa}

\usepackage{graphicx}
\usepackage{epstopdf}
\usepackage{txfonts}
\usepackage{lscape}

\begin{document} 

\title{Reaching the boundary between stellar kinematic groups \\ and very wide binaries}
\subtitle{III. Sixteen new stars and eight new wide systems in the $\beta$~Pictoris moving group}
\titlerunning{Sixteen new stars and eight new wide systems in the $\beta$~Pictoris moving group}
%
%\authorrunning{...}

 %
  \author{F. J. Alonso-Floriano\inst{1}
          \and
          J. A. Caballero\inst{2}
         \and
         M. Cort{\'e}s-Contreras\inst{1}
          \and
         E. Solano\inst{2,3}
         \and
         D. Montes\inst{1}
          }
\institute{
        Departamento de Astrof{\'i}sica y Ciencias de la Atm{\'o}sfera, Facultad de Ciencias F\'{\i}sicas, Universidad Complutense de Madrid, 28040 Madrid, Spain \\ \email{fjalonso@ucm.es}
         \and
        Centro de Astrobiolog\'ia (CSIC-INTA), ESAC PO box 78, 28691 Villanueva de la Ca\~nada, Madrid, Spain
        \and
        Spanish Virtual Observatory, ESAC PO box 78, 28691 Villanueva de la Ca\~nada, Madrid, Spain
         }
\date{Received 19 June 2015 /  Accepted {8} August 2015}

% \abstract{}{}{}{}{} 
% 5 {} token are mandatory
 
  \abstract
   {}
   {We look for common proper motion companions to stars of the nearby young $\beta$\,Pictoris moving group.}
   {First, we compiled a list of 185 $\beta$\,Pictoris members and candidate members from 35 representative works.
Next, we used the Aladin and STILTS virtual observatory tools and the PPMXL proper motion and Washington Double Star catalogues to look for companion candidates.
The resulting potential companions were {subjects} of a dedicated astro-photometric follow-up using public data from all-sky surveys.
After discarding 67 sources by proper motion and 31 by colour-magnitude diagrams, we obtained a final list of 36 common proper motion systems.
The binding energy of two of them is perhaps too small to be considered physically bound.}
   {Of the 36 pairs and multiple systems, eight are new, 16 have only one stellar component previously classified as a $\beta$\,Pictoris member, and three have secondaries at or below the hydrogen-burning limit.
Sixteen stars are reported here for the first time as moving group members.
The unexpected large number of high-order multiple systems, 12 triples and two quadruples among 36 systems, may suggest a biased list of members towards close binaries {or an increment of the high-order-multiple fraction for very wide systems.}}
{}
\maketitle

%
%________________________________________________________________

\section{Introduction}

\begin{table*}[]
\centering
\caption{Sources of the $\beta$ Pictoris stellar sample.}
\label{tab.samplesources}
\begin{tabular}{l l} 
   \hline
   \hline
   \noalign{\smallskip}
Title                                    & References  \\
  \noalign{\smallskip}
    \hline
\noalign{\smallskip}
Search for associations containing young stars (I, III, V, VI)  &       SACY\,\tablefootmark{a}                                 \\
Bayesian analysis to identify new star candidates in nearby young stellar...    (I--V)& BANYAN\,\tablefootmark{b}       \\
A dusty M5 binary in the $\beta$\,Pictoris moving group &       Rodr{\'i}guez et~al. 2014\\
On the age of the $\beta$\,Pictoris moving group                &       Mamajek \& Bell 2014    \\
The Solar Neighborhood. XXXIII. Parallax results from the CTIOPI\,0.9\,m program...                                              &       Riedel et~al. 2014\\
A lithium depletion boundary age of 21 Myr for $\beta$\,Pictoris moving group   &       Binks \& Jeffries 2014        \\
Unveiling new members in five nearby young moving groups & Mo{\'o}r et~al. 2013\\
Identifying the young low-mass stars within 25 pc (I, II)&  Shkolnik et~al. 2009, 2012      \\
Likely memebers of the $\beta$\,Pictoris and AB Doradus moving groups in the north & Schlieder et~al. 2012b      \\
Cool young stars in the northern hemisphere: $\beta$\,Pictoris and AB Doradus moving... &      Schlieder et~al. 2012a         \\
The sizes of the nearest young stars                            &       McCarthy \& White 2012           \\
Potential members of stellar kinematic groups within 30\,pc of the Sun  &       Nakajima \& Morino 2012          \\
A search for new members of the $\beta$\,Pictoris, Tucana-Horologium and $\eta$\,Cha& Kiss et~al. 2011                           \\
$\beta$\,Pictoris and AB\,Doradus moving groups: likely new low-mass members & Schlieder et~al. 2010 \\
The lowest-mass member of the $\beta$\,Pictoris moving group                 &  Rice et~al. 2010 \\
Potential members of stellar kinematic groups within 20\,pc of the Sun  & Nakajima et~al. 2010            \\
Kinematic analysis and membership status of TWA22 AB            &Teixeira et~al. 2009 \\
Nearby young stars selected by proper motion. I. Four new members of the $\beta$\,Pictoris...    & L{\'e}pine \& Simon 2009              \\
Young nearby loose associations                                         & Torres et~al. 2008                      \\
Unraveling the origins of nearby young stars            & Makarov 2007                                  \\
Nearby debris disk systems with high fractional luminosity reconsidered & Mo{\'o}r et~al. 2006 \\
Young stars near the Sun                                                        & Zuckerman \& Song 2004          \\
New aspects of the formation of the $\beta$\,Pictoris moving group      & Ortega et~al. 2004      \\
New memeber of the TW Hydrae association, $\beta$\,Pictoris moving group and Tucana... & Song et~al. 2003\\
The origin of the $\beta$\,Pictoris moving group                & Ortega et~al. 2002     \\
The $\beta$\,Pictoris moving group                                      & Zuckerman et~al. 2001b\\
The age of $\beta$\,Pictoris                                            & Barrado y Navascu{\'e}s et~al. 1999     \\

\noalign{\smallskip}
\hline
\end{tabular}
\tablefoot{
\tablefoottext{a}{SACY: Torres et~al. (2006); da Silva et~al. (2009); Elliott et~al. (2014, 2015).}
\tablefoottext{b}{BANYAN: Malo et~al. (2013, 2014a, 2014b); Gagn{\'e} et~al. (2014, 2015). 
We only collected BANYAN candidates with membership probability $P$\,>\,50\%.}}
\end{table*}

Wide binaries provide valuable information about key questions in astrophysics; for example, halo-wide pairs contribute to constraining the properties of dark matter (Weinberg et~al. 1987; Yoo et~al. 2004; Quinn et~al. 2009), some star formation theories depend on the frequency and separation of wide young binaries (Parker et~al. 2009; Ward-Duong et~al. 2015; Marks et~al. 2015), and relatively bright FGK-type primaries with M-dwarf companions provide a metallicity calibration yardstick for cool stars (Bonfils et~al. 2005; Rojas-Ayala et~al. 2012; Newton et~al. 2014; Li et~al. 2014).
However, the maximum projected physical separation of a wide binary is still a matter of discussion:
Some authors consider a cutoff in the number of wide binaries at 2\,10$^{4}$\,au ($\sim$0.1\,pc), which is the typical size of protostellar cores (Tolbert 1964; Abt 1988; Wasserman \& Weinberg 1991; Allen et~al. 2000; Tokovinin \& L{\'e}pine 2012), while others contemplate separations of 2\,10$^{5}$\,au ($\sim$1\,pc) or more (Jiang \& Tremaine 2009; Caballero 2009; Shaya \& Olling 2011).
Such wide common proper-motion pair candidates, which give their name to the title of this series of papers, can be either unbound members of the same young stellar kinematic group that by chance are co-moving (Tokovinin 2014a) or bound ``binaries'' of very low binding energies at the limit of disruption (Caballero 2010).

The younger a weakly bound system is, the less time it has had to be disrupted (Bahcall \& Soneira 1981; Retterer \& King 1982; Weinberg et~al. 1987; Saarinen \& Gilmore 1989; Poveda \& Allen 2004).
As a result, a search for multiple systems within a young stellar kinematic group (moving group or stellar association) offers a unique opportunity for finding new faint benchmark objects hardly influenced by the Galactic gravitational potential, but instead by their formation process.
In other words, the shape of young wide binaries is dominated by {\em \emph{nature}} instead of by {\em \emph{nurture}}.

In this work, we use the profitable technique of searching for common proper-motion pairs of wide separation (e.g., Luyten 1979; Chanam{\'e} \& Gould 2004) in a close and very young moving group, namely $\beta$\,Pictoris (Zuckerman et~al. 2001b; Ortega et~al. 2002; Song et~al. 2003).
Although there is no consensus in the literature, the $\beta$\,Pictoris age lies in a relatively narrow interval between 11\,Ma and 26\,Ma (Barrado y Navascu\'es 1998; Torres et~al. 2006; Yee \& Jensen 2010; Binks \& Jeffries 2014; Mamajek \& Bell 2014 and references therein).
Known moving group members and member candidates lie at between 6\,pc and 80\,pc from our Sun with a median distance of 40\,pc.

Because of its youth and proximity, the $\beta$\,Pictoris moving group has been relevant for studying resolved debris discs with high angular resolution observations (Smith \& Terrile 1984; Metchev et~al. 2005; Boccaletti et~al. 2009; Churcher et~al. 2011; Wahhaj et~al. 2013; Dent et~al. 2013) and exoplanets through direct imaging (Mouillet et~al. 1997; Neuh{\"a}user et~al. 2003; Kasper et~al. 2007; Lagrange et~al. 2009, 2010; Bonnefoy et~al. 2011, 2013; Biller et~al. 2013; Rameau et~al. 2013; Males et~al. 2014; Bowler et~al. 2015; {Macintosh et~al. 2015}) or for comparing observations with evolutionary models (Crifo et~al. 1997; Song et~al. 2002; Cruz et~al. 2009; Biller et~al. 2010; Mugrauer et~al. 2010; Jenkins et~al. 2012; {Montet et~al. 2015}).
Therefore, increasing the number of members via common proper-motion companionship, especially at low masses, can help {to inform} the previously mentioned fields and to constrain the age of the group.
{Besides that, identiying bright M-dwarf targets of $\sim$10--30\,Ma for extremely precise radial velocity surveys is becoming critical for understanding the formation and early evolution of terrestrial planets in habitable zones (Lissauer 2007; Ram{\'i}rez \& Kaltenegger 2014; Luger et~al. 2015; Tian 2015; Tian \& Ida 2015).}
Preliminary results of this work, including the discovery of two new stellar members in the $\beta$\,Pictoris moving group, were given in Alonso-Floriano et~al.~(2011).

%
%________________________________________________________________
\section{Analysis\label{sec.analysis}}
\subsection{Stars sample\label{sec.sample}}

We have compiled in Table~\ref{tab.sample} a list of 185 $\beta$\,Pictoris members and member candidates around which we looked for common proper-motion companions.
We gathered them from 35 previous works published in the past 16 years from the first articles of Barrado y Navascu{\'e}s et~al. (1999) and Zuckerman et~al. (2001b) to the last investigations published in the SACY (Search for Associations Containing Young stars -- Torres et~al. 2006; Elliott et~al. 2014, 2015) and BANYAN series (Bayesian Analysis for Nearby Young AssociatioNs -- Malo et~al. 2014a,b; Gagn{\'e} et~al. 2015).
Table~\ref{tab.samplesources} lists all works that we searched through.

We cross-matched our list with the latest Geneva-Copenhagen catalogue (Holmberg et~al. 2009) and identified 17 bright stars for which metallicity was available.
From these data, we determined a solar metallicity of the $\beta$\,Pictoris moving group {of} [Fe/H]\,=\,--0.2$\pm$0.2. In Table~\ref{tab.sample}, we provide for each star:
discovery (or recommended) name, right ascension and declination from the Two-Micron All-Sky Survey (Skrutskie et~al. 2006), heliocentric distance, its uncertainty when available, and corresponding reference.
We follow the nomenclature convention of Alonso-Floriano et~al. (2015).
In particular, we provide for the first time the {\em ROSAT} precovery names (1RXS) for several stars for which no X-ray counterpart had been identified by subsequent proper-motion surveys.
 
In the last column, we also list a flag indicating the quality of the star membership in $\beta$\,Pictoris:
\begin{enumerate}
\item Uncontrovertible moving group members for which at least two independent research groups have declared them to be {\em \emph{bona fide}} moving group members and whose {memberships have not been} put in doubt afterwards.
In general, these objects have coherent kinematics (with reliable distance and radial velocity determination) and youth features (coronal X-ray and chromospheric H$\alpha$ emission, lithium in absorption and, in some cases, debris discs).
\item Moving-group member candidates for which there is no definitive confirmation of true membership.
\item Dubious moving group member candidates that have also been proposed as belonging to other young moving groups of similar kinematics, or even to the field.
We include them in our work for completeness.
\end{enumerate}

%
%________________________________________________________________

\subsection{Proper motion companion candidates\label{sec.pm_comp_cand}}

\begin{figure}[]
\includegraphics[trim=70 20 100 20, width=0.49\textwidth]{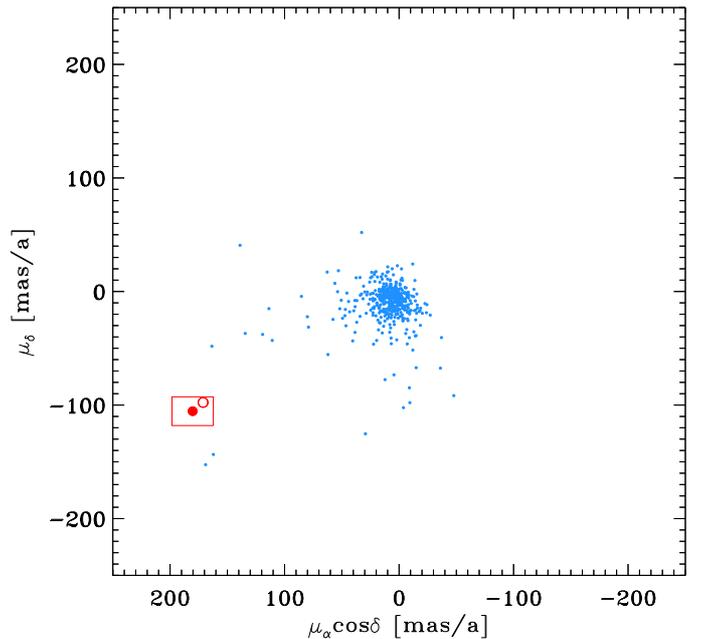}%10 20 25 20
 \caption{\label{fig.PPMXL_example} Representative proper-motion diagram of all PPMXL sources brighter than $J$ = 15.5\,mag in a 30\,arcmin-radius circular area centred on \object{LP~648--20}.
The red square box in the bottom left indicates the proper-motion search area around LP~648--20, marked with a filled circle.
The open circle corresponds to the bright, young G5\,V star \object{EX~Cet}.}
\end{figure}

For this search, we made extensive use of virtual observatory tools. 
We used the comprehensive PPMXL proper motion catalogue (Roeser et~al. 2010), the Aladin sky atlas (Bonnarel et~al. 2000), and the Starlink Tables Infrastructure Library Tool Set (STILTS; Taylor 2006) to look for common proper-motion companions to the 185 $\beta$\,Pictoris stars in Table~\ref{tab.sample}.
The PPMXL catalogue is complete down to the visual magnitude $V$\,$\approx$\,20\,mag and has typical individual mean errors of the proper motions between 4 and 10\,mas/a, approximately.
We applied the following selection criteria in our search.
\begin{itemize}

\item We looked for companion candidates in a circular area of angular radius $\rho$\,=\,$s / d$ (in arcsec) centred on each sample star, where $s$ is the maximum projected physical separation, fixed at $s$\,=\,10$^5$\,au, and $d$ (in pc) is the heliocentric distance shown in Table~\ref{tab.sample}.
At the given distances, the search radii varied between over 4\,deg for the closest $\beta$\,Pictoris stars (e.g., \object{YZ~CMi}~AB at 5.96$\pm$0.08\,pc) and {12 to 23\,arcmin} for the most distant ones {(e.g., \object{LP~58--170} at 140$\pm$40\,pc and \object{V4046 Sgr}~AB and~C at 73$\pm$18\,pc)}.
The median search radius was 44\,arcmin.

\item We discarded from the survey 24 stars with total PPMXL proper motions $\mu$\,<\,50\,mas/a (19) or no proper motions at all (5). 
Therefore, we looked for companions of 161 $\beta$\,Pictoris stars.
Stars slower than $\mu$\,=\,50\,mas/a were not considered at this step because of the large number of potential candidates with relative uncertainties of 10\,\%--30\,\% in proper motion that would {fall} in the surveyed area and pass the filter. As proper motion companion candidates,
we classified only the PPMXL sources with a 2MASS counterpart for which the values of $\mu_\alpha \cos{\delta}$ and $\mu_{\delta}$ lie within 10\,\% of those of the primary target (see Fig.~\ref{fig.PPMXL_example}).

\item We retained objects brighter than $J$\,=\,15.5\,mag.
In general, fainter sources in the near-infrared also have very faint magnitudes in the optical, close to the limit of the USNO-B1 (Monet et~al. 2003) digitisations of $B_J$, $R_F$, and $I_N$ photographic plates, which were used by PPMXL.
This faintness translates into large astrometric errors in the PPMXL proper motions.
Keeping relatively bright sources assures the quality of the compiled astro-photometric measurements (see below), although prevents detecting fainter and, thus, low-mass $\beta$\,Pictoris members in, perhaps, the substellar domain.

\end{itemize}

Once we had a preliminary list of candidates, we {inspected all of them visually} with Aladin and the images and data of various all-sky surveys (Palomar Observatory Sky Survey I and II; 2MASS; SDSS-DR9, Ahn et~al. 2012; {\em WISE}, Cutri et~al. 2012, 2014; CMC14 and CMC15, Evans et~al. 2002).
In particular, we checked that the companion candidates have a unique and reliable entry in the PPMXL catalogue (e.g., at least four astrometric detections, no other PPMXL source at less than 2\,arcsec, smooth variation of the magnitudes from $B_J$, through $R_J$, $I_N$, $J$, $H$, $K_s$, to {\em WISE} $W$1--4).
In this step, we discarded a number of preliminary companion candidates with erroneous PPMXL proper motions (i.e., with incorrect USNO-B1 matches) owing to close visual multiplicity or source confusion in very crowded fields at low Galactic latitudes.
Some of the mistaken sources were identified around
\object{$\alpha$~Cir},                  % b=-04.5917
1RXS J171502.4--333344,
\object{V4046~Sgr},                     % b=-07.2301
\object{1RXS~J184956.1--013402},        % b=-00.3787
which have $|b|$\,$<$\,7\,deg,
and, especially, \object{V343~Nor},     % b=-01.8024
which is at less than 2\,deg of the Galactic plane and, besides this, towards the Galactic centre.
After this visual pre-cleaning, we obtained a list of 92 proper motion companion candidates to 65 $\beta$\,Pictoris stars.

Next, we performed a 10\,arcsec-radius cross-match on our initial 185-star sample  with the Washington Double Star catalogue (WDS -- Mason et~al. 2001, 2015).
We got 163 positive cross-matches in 55 WDS systems.
Of the cross-matches, 136 corresponded to close physical binaries not resolved by 2MASS nor PPMXL ($\rho$\,$\lesssim$\,2.5\,arcsec) or to wider multiple systems, but with large magnitude differences measured with powerful adaptive optics systems (e.g., Lafreni\`ere et~al. 2007; Chauvin et~al. 2010).
The list of WDS systems unresolved or unidentified in our search are shown in Table~\ref{tab.WDS}, which provides the star name, WDS discovery name (for resolved pairs) or reference (for spectroscopic binaries), multiplicity status (physical, visual {--non-common proper motion--}, single/double-line spectroscopic binaries), angular separation (interval of $\rho$ when several visual companions are tabulated), position angle {($\theta$)}, and WDS identifier (only for resolved pairs).
{For the physical and visual pairs in Table~\ref{tab.WDS}, $\rho$ and $\theta$ correspond to the latest epoch listed by WDS.}  

Thanks to the cross-match with WDS, we were able to add another 15 previously known secondaries detected by 2MASS to our list of 92 proper motion companion candidates.
They did not pass our filters above because they have PPMXL proper motions that deviate more than 10\,\% from those of the ``primary'', probably because of relative orbital motion, erroneous measurements in right ascension and/or declination, {proper motions with 1$\sigma$ lower limits below} the 50\,mas/a boundary, or no proper motions at all.

 \begin{figure*}[]
 \centering
\includegraphics[trim=70 20 100 20, width=0.49\textwidth]{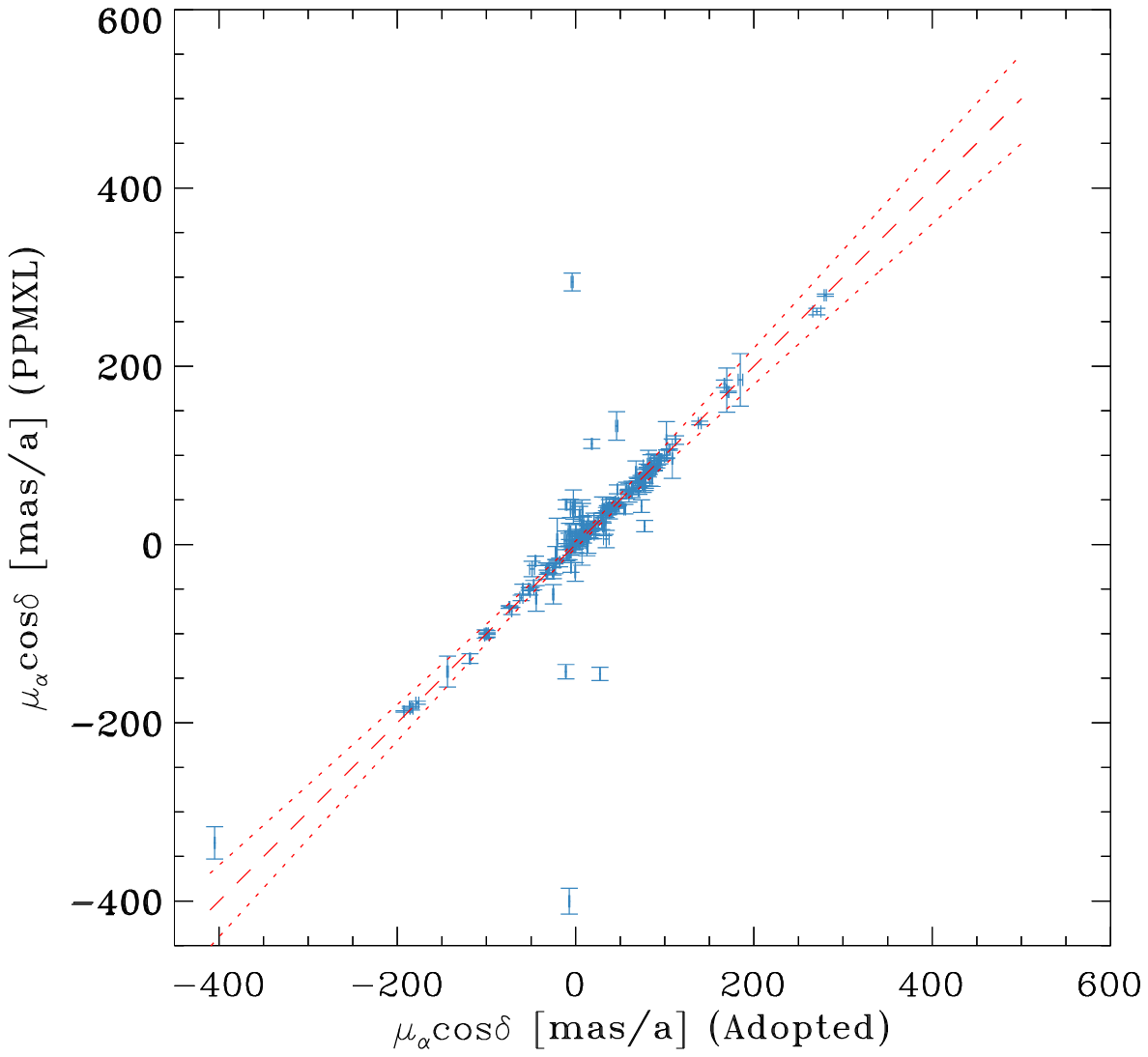}
\includegraphics[trim=70 20 100 20, width=0.49\textwidth]{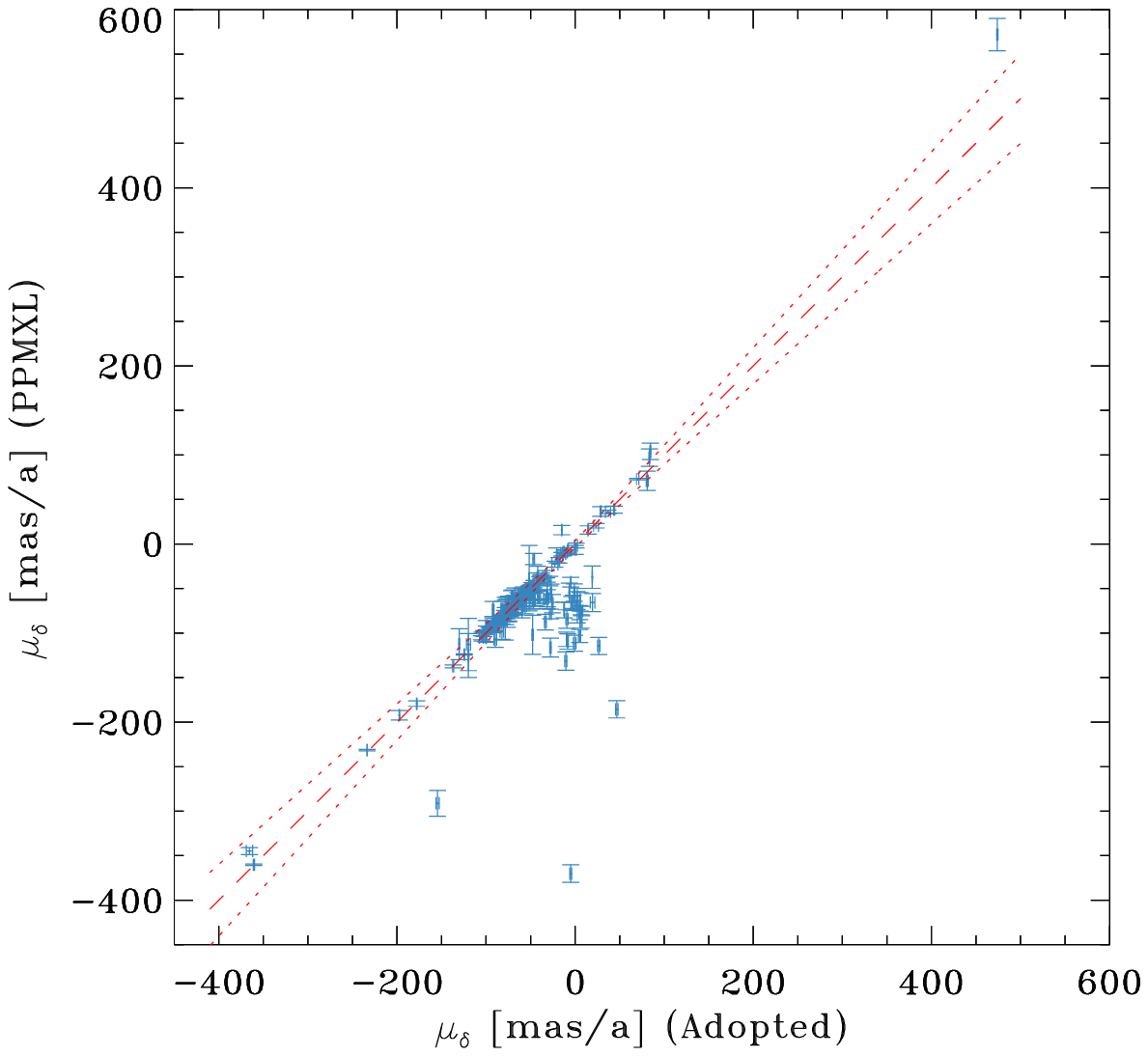}
 \caption{\label{fig.PPMXLvsAdopted} PPMXL vs. adopted proper-motion diagrams in right ascension ({\em left panel}) and declination ({\em right panel}).
The red dashed and dotted lines mark the one-to-one relationship and the 10\,\% error area above and below it, respectively.
} 
\end{figure*}

Finally, we also added to our list the companion candidates of three additional pairs of $\beta$\,Pictoris stars that {were} not in WDS and that were not detected because of the reasons explained above:
{[SLS2012]}\,PYC\,J02017+0117N \& S ($\rho \sim$ 10\,arcsec and equal brightness) and TYC\,112--917--1 \& 2E\,1249\,AB ($\mu$\,$\approx$\,41\,mas/a), which have the same predicted or measured distances and radial velocities (Schlieder et~al. 2012a; Elliott et~al. 2014) and are quite obvious proper motion companion candidates in Aladin, and V4046~Sgr~AB and~C, which was presented by Kastner et~al. (2011).

In Table~\ref{tab.astrometry}, we list the 110 (92+15+3) proper motion companion candidates that passed on to the next analysis stage.

%
%________________________________________________________________
\subsection{Astro-photometric follow-up\label{sec.follow-up}}

\subsubsection{Astrometry}

We performed a dedicated astro-photometric follow-up of the 110 companion candidates in two steps.
In the first one, we confirmed true common proper motion of the pairs with a precise astrometric study.
This step was necessary because PPMXL used the astro-photometric USNO-B1 catalogue as input, which is known to be affected by systematics at the fainter optical magnitudes, especially when dealing with high proper motion stars.

Of the 184 objects in Table~\ref{tab.astrometry} (74 primaries and 110 companion candidates), 55 had reliable proper motions measured by {\em Hipparcos} (TYC, H{\o}g et~al. 2000; HIP2, van~Leeuwen 2007).
For the remaining 129 objects, we measured precise proper motions from public data in virtual observatory catalogues as in Caballero (2010, 2012).
In particular, we used astrometric epochs from the following catalogues:
AC2000.2 (Urban et~al. 1998),
USNO-A2 (Monet 1998),
GSC2.3 (Lasker et~al. 2008),
DENIS (Epchtein et~al. 1997),
CMC14 and CMC15,
2MASS,
SDSS, and
{\em WISE}.
To maximise the number of astrometric epochs, $N$, and time baseline, $\Delta t$, of the follow-up, which translates into reducing the uncertainty in proper motion, we also used the SuperCOSMOS digitisations of Palomar Observatory Sky Survey photographic plates, especially for the faintest objects (Hambly et~al. 2001; cf., Caballero 2012).
The addition of SuperCOSMOS data allowed us to get at least four accurate astrometric epochs spread over a minimum of 11.5\,a for all targets except for one star (2MASS J05113065--2155189, $N$ = 3).
{The average} number of astrometric epochs and time baseline were five and 34\,a, respectively.
In the extreme case of \object{TYC~4571--1414--1}, we measured its proper motion with eight astrometric epochs spread over almost 115\,a.
Table~\ref{tab.astrometry} lists the 2MASS coordinates of the 184 "primaries" and companion candidates, and their PPMXL and adopted proper motions.
For the adopted proper motions that do not come from TYC or HIP2, Table~\ref{tab.astrometry} also {provides} the time baseline and number of epochs used in our astrometric follow-up.
Except for partially resolved close binaries (e.g., \object{AT~Mic}~AB) or faint sources ($r'$\,$\gtrsim$\,16\,mag), we were able to measure proper motions with typical uncertainties of 1\,mas\,a$^{-1}$ or less, which are comparable to or even better than TYC or HIP2.

We show a comparison of the original PPMXL proper motion values and the ones adopted by us in Fig.~\ref{fig.PPMXLvsAdopted}.
While the values of proper motions in right ascension provided by PPMXL have in general good agreement with our adopted values, many PPMXL proper motions in declination have greater absolute values.

With the new data, we made a second, more precise, astrometric filtering and discarded 43 visual companions with adopted proper motions that deviate more than 10\,\% from the proper motion of the system.
This step of the follow-up thus left 67 physical companion candidates for the second step.
In general, the astrometrically rejected objects are distant background stars with fake high proper motions in the PPMXL catalogue, which are located close to bright stars and were reported previously as companion candidates or which are located at large angular separations to our primary targets and have by chance similar, but not identical, proper motions.
Of the rejected stars, four were catalogued companion candidates of \object{c~Eri}, \object{$\alpha$~Cir}, \object{CD--24~16238}, and \object{AF~Psc}, and one was a faint companion candidate to \object{[SLS2012] PYC~J10175+5542} (Schlieder et~al. 2012a) that had been reported by W.\,J. Luyten (LDS~2851, WDS~10176+5542).

%________________________________________________________________
\subsubsection{Photometry}

 \begin{figure*}[]
 \includegraphics[trim=15 10 35 40, width=0.49\textwidth]{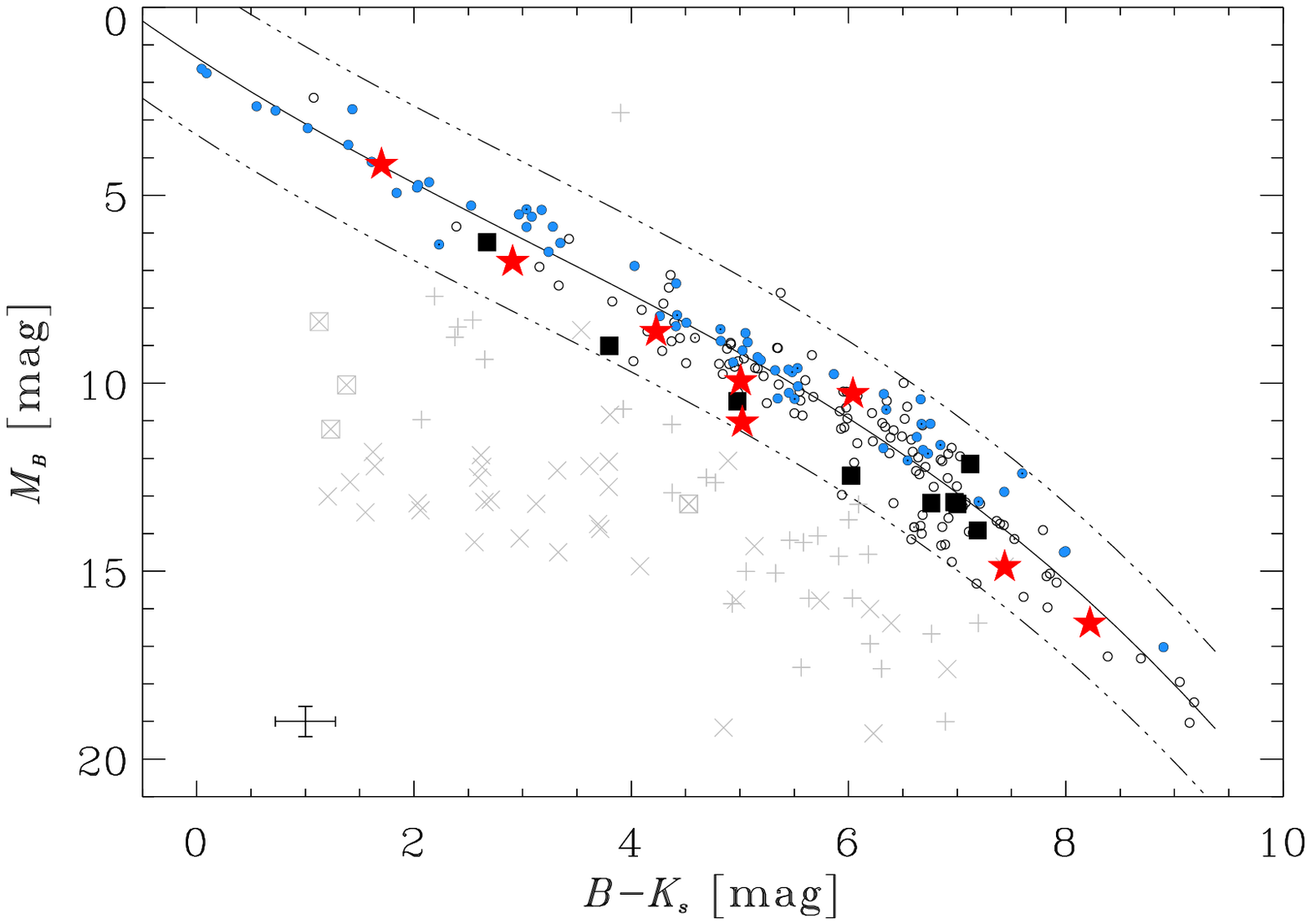}
\includegraphics[trim=10 10 40 40, width=0.49\textwidth]{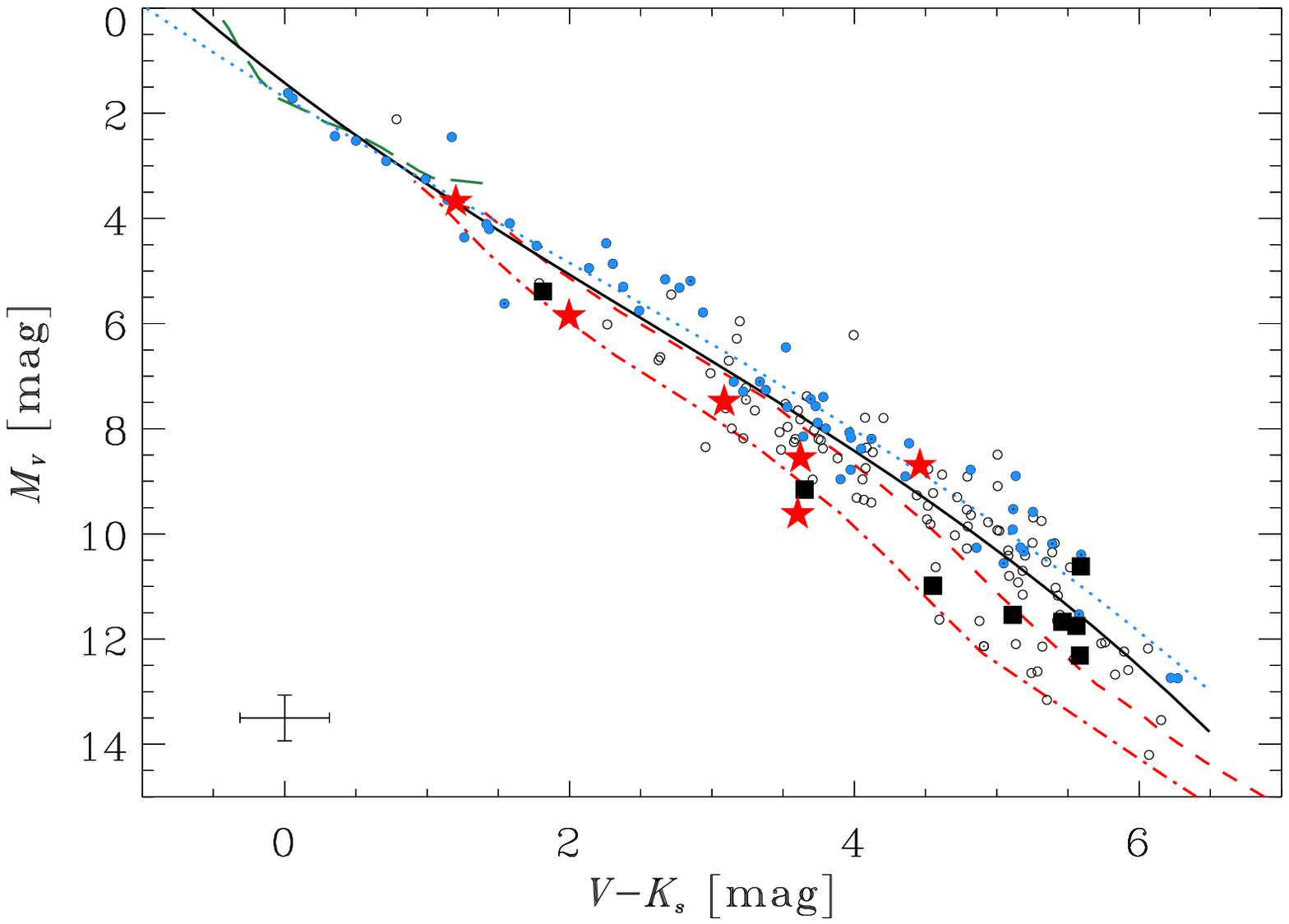}
 \caption{\label{fig.VK_models}  Optical-to-near-infrared colour-magnitude diagrams of our $\beta$~Pictoris stars and common proper-motion companion candidates.
In both panels, blue filled circles mark {\em \emph{bona fide}} moving group members with membership flag 1 in Table~\ref{tab.sample} (see Section 2.1), and open circles indicate other member candidates with flags 2 and 3.
Black filled squares denote companions previously known in the literature that had not been reported as belonging to $\beta$\,Pictoris.
Red filled stars mark our eight new proper motion companions.
Typical error bars are shown in the bottom left corner.
The black solid line is the average $\beta$~Pictoris sequence computed with all candidate members as in the left panel (flags 1, 2, and 3).
{\em Left panel}: $M_B$ vs. $B-K_s$ diagram.
The dash-dotted lines are the $\beta$~Pictoris sequence shifted by $\pm 3 \sigma$.
Grey times ($\times$) and crosses ($+$) indicate discarded companion candidates from astrometry and photometry, respectively; squared symbols mark companion candidates in the WDS.
{\em Right panel}: $M_V$ vs. $V-K_s$ diagram.
The blue dotted line is the sequence with only {\em \emph{bona fide}} members (flag 1).
Red dashed and dash-dotted lines are the 20 and 100\,Ma isochrones from Baraffe et~al. (2015).
Green long dashed line is the 20\,Ma isochrone from Siess et~al. (2000) plotted only at highest masses.
For clarity, we do not draw the discarded companion candidates.
Some remarkable stars do not have $V$ photometry.
\label{fig.BK}}
\end{figure*}

In the second step of the follow-up, we studied the membership in the $\beta$\,Pictoris moving group of the 67 companion candidates that passed the previous astrometric filter with the help of colour-magnitude diagrams and theoretical isochrones.

First, we compiled $B$, $V$, $r'$, $J$, $H$, $K_s$, and $W$1--4 magnitudes for {\em \emph{all}} the sources investigated in this work.
While infrared $JHK_sW$1--4 photometry came in all cases from 2MASS and {\em WISE}, the origin of the optical $BVr'$ photometry was diverse.
When available, we collected $BVr'$ photometry from UCAC4 (Zacharias et~al. 2013).
If not available, we got it from a number of sources:
Tycho-2 ($B$ and $V$, after transformation from $B_T$ and $V_T$ magnitudes),
USNO-B1 ($B$, after average of two photographic $B_J$ magnitudes),
AC2000.2 (only one star),
SPM4 (Girard et~al. 2011 -- $V$, only one star),
CMC 15 ($r'$),
SDSS DS9 ($r'$),
or the literature (Voges et~al. 1999; Bakos et~al. 2002; Torres et~al. 2006; Beichman et~al. 2010; Smart 2013).

{We were not able to compile optical $BVr'$ photometry for all our targets. 
Since we were able to compile magnitudes for more stars in the $B$ band, we applied our photometric filtering using the reddest near-infrared band, $K_s$, and the bluest optical one, $B$.
Actually, we failed to find reliable $B$ photometry for only six stars:
five low-mass stars or brown-dwarf candidates with spectral types at the M/L boundary, of which four are from Gagn{\'e} et~al. (2014, 2015), one is the known companion of L~186--67~A (see below), and the sixth one is a star close to the bright primary V343~Nor~A. 
The reddest and faintest object in our sample with $B$ and $K_s$ photometry is \object{2MASS~J06085283-2753583} (M8.5\,V, Luhman et~al. 2009; $B-K_s$\,=\,9.1\,mag).
The use of {\em WISE} photometry did not provide significant improvement over the use of 2MASS $K_s$.}

We performed our photometric filtering in a recursive scheme:

\begin{itemize}

\item First we computed the $B$-band absolute magnitude $M_B$ with the heliocentric distances in Table~\ref{tab.sample} and built the $M_B$ vs. $B-K_s$ diagram in left-hand panel of Fig. 3.
Any new companion candidate {would have to be located at the same distance as the target star}.

\item We defined an average $\beta$~Pictoris sequence with all members and candidates in Table~\ref{tab.sample} (flags 1, 2, and 3).
All sources that did not pass the astrometric filter lie in the locus of background stars in the colour-magnitude diagram.
\item We checked the reliability of our average sequence by comparing it with the latest evolutionary models by Baraffe et~al. (2015).
Since BT-Settl does not provide $M_B$ magnitudes, we used $M_V$ ones instead.
We built the $M_V$ vs. $V-K_s$ diagram in right-hand panel of Fig. 3 and plotted the corresponding BT-Settl 20 and 100\,Ma isochrones.
The acceptable match between our average $M_V$ vs. $V-K_s$ sequence and the 20\,Ma isochrone encouraged us to use our average $M_B$ vs. $B-K_s$ sequence for the photometric filter.

\item We discarded stars with absolute magnitudes $M_B$ and colours $B-K_s$ inconsistent with the $\beta$\,Pictoris sequence.
Most discarded stars lie outside the $\pm 3 \sigma$ area around the sequence.
The systematic error introduced by mixing different photometric systems for the blue magnitude seems to be smaller than the intrinsic scatter in the $\beta$\,Pictoris sequence, mostly due to uncertainties in distance.

\end{itemize}

Of the previous 67 stars, 31 did not pass the photometric filter.
One known system, L~186--67~Aa,Ab,B, could not be studied photometrically because of the lack of reliable data in the optical, but the short angular separation between components and the large common proper motions ensured that it is a physical system.

%
%________________________________________________________________
\section{Results and discussion\label{sec.results}}

\subsection{Known and new common proper motion pairs}

From the initial list of 184 common proper motion companion candidates to $\beta$~Pictoris stars in Section~\ref{sec.pm_comp_cand}, only 36 targets passed the two filters of our astro-photometric follow-up in Section~\ref{sec.follow-up}. 
Our final list of confirmed common-proper motion systems in the $\beta$~Pictoris moving group, as shown in Table~\ref{tab.results}, consists of

\begin{itemize}

\item Eighteen known systems in which the two stars had been reported previously to belong to the moving group.
All of them except one are listed by WDS; the exception is the wide system formed by \object{V4046~Sgr}~AB and V4046~Sgr~C, which was proposed and investigated by the first time by Kastner et~al. (2011).
Some of the 17 WDS systems have been known for decades, such as five pairs in the W.\,J. Luyten's Double Star catalogue or the \object{HD~14082}~AB pair, which was resolved for the first time by F.\,G.\,W. Struve in 1821.

\item Ten known systems in which only {\em \emph{one}} star had been reported previously as belonging to the moving group.
Of the ten stars that had not been reported as belonging to $\beta$~Pictoris (i.e., not listed in Table~\ref{tab.sample}), four displayed significant X-ray emission in {\em ROSAT} observations (Voges et~al. 1999; Riaz et~al. 2006; Kaplan et~al. 2006; Haakonsen \& Rutledge 2009) and two showed intense H$\alpha$ emission at the chromospheric/accretion boundary for their spectral types ($pEW$(H$\alpha$) $\approx$ --12 to --16\,{\AA} -- Reid et~al. 1995; Riaz et~al. 2006).
Since there is one star that is an H$\alpha$ {\em and} X-ray emitter (\object{2MASS~J00193931+1951050}), half of the ten new stars have known signposts of youth, which supports membership in $\beta$~Pictoris.
Besides this, another one has a similar radial velocity to the primary in the system (\object{CD--44~753}~A and B -- Kordopatis et~al. 2013).
For the other four new young stars, there are only photometric data available (and, in the case of \object{2MASS~J07293670+3554531}, mass and spectral type derived from photometry -- Pickles \& Depagne 2010; Janson et~al. 2012).

\item Eight new common proper motion systems with $\beta$~Pictoris stars.
In reality, there are WDS entries for two $\beta$~Pictoris pairs that were presented for the first time by Alonso-Floriano et~al. (2011): \object{EX~Cet}~A,B (CAB~3) and \object{HD~173167}~A,B (CAB~8).
Although the results from this preliminary publication have already been used by other authors (Shkolnik et~al. 2012; Eisenbeiss et~al. 2013; Bowler et~al. 2015), we consider their discovery as part of this work.
Mo\'or et~al. (2003) ``rediscovered'' the pair {HD~173167}~A,B, although they did not report $\rho$ or $\theta$.
The optical spectra of these two stars and of {TYC~112--917--1} and {2E~1249}~AB in the new pair {\em WDS~05200+0613} display intense Li~{\sc i} $\lambda$6707.8\,{\AA} line in absorption for their spectral type (Alcal\'a et~al. 2000; Torres et~al. 2006), which supports their {extremely} young age. Six of these stars are reported here as new member candidates in the $\beta$~Pictoris moving group.

\end{itemize}

On some occasions we use the term ``pair'' to refer to multiple systems that contain only two components resolvable from the ground with standard imaging (i.e., no adaptive optics or lucky imaging) and spectroscopic devices. 
Most of our systems are such pairs.
However, Table~\ref{tab.results} lists 12 triple and two quadruple hierarchical systems that contain one or two close pairs unresolved by public catalogues (Table~\ref{tab.WDS}).
The two quadruple systems are \object{MV~Vir}~Aa,Ab,B,C and HD~199143 AB,CD ({for which the} close components were resolved first by Jayawardhana \& Brandeker 2001).
The latter has an {"A.\,Tokovinin"} WDS entry dated after 2011, but the wide multiplicity was previously reported by Alonso-Floriano et~al. (2011) and, especially, Zuckerman et~al. (2001b).

{The existence of 14 triples and quadruples in a list of 36 multiples provides a high-order-multiple ratio of about 1:3, which is unexpectedly high.
Law et~al. (2010) found a similar ratio of about 1:2 for wide M-dwarf binaries of the field and suggest that some of the binaries with large separations are actually triple and quadruple systems. (Actually, Caballero 2007 and Burgasser et al. 2007 pointed it out before.)
The increment of the high-order-multiple fraction for the widest systems is supported by the work of Reipurth \& Mikkola (2012), who used $N$-body simulations of the dynamical evolution of triple systems to suggest that loosely bound triple systems might appear to be very wide binaries.
However, recent dedicated surveys for multiplicity of F, G, K (Tokovinin et~al. 2014b; Elliott et~al. 2015)  and M dwarfs in the field (Cort{\'e}s-Contreras et~al. 2014) have found lower ratios of about 1:10.
Although our sample comprises a wide range of masses and separations, it is not large enough to do an appropriate comparison with the previously mentioned works.
Another explanation might be an observational effect of a biased sample in which surveys for nearby young stars are naturally slanted towards detecting intrinsically bright binaries and multiple stars (Malmquist bias), and active spectroscopic binaries (very close separations enhances stellar activity).
The discovery of new moving groups members only based on astrometry, as in this survey, may help to alleviate this observational bias.}

In Table~\ref{tab.results} we list our WDS identifiers in italics if they are not included in the WDS catalogue at the time of writing these lines (i.e., V4046~Sgr~AB,C and six of the eight new pairs).
In total, in this survey we propose 16 new stellar members of the $\beta$~Pictoris moving group: six in new pairs and ten in known systems with only one reported young star.
One of the new $\beta$~Pictoris stars in a new pair is \object{HD~173167}~A, which was discovered by Alonso-Floriano et~al. (2011) and classified afterwards as a moving group member by Mo\'or et~al. (2013). 
These values represent an increase of 9\,\% in the total number of reported $\beta$~Pictoris stars and of almost 30\,\% in the number of wide proper motion systems in the moving group.
{We ran the on-line BANYAN tool\footnote{http://www.astro.umontreal.ca/$\sim$malo/banyan.php} (Malo et~al. 2013) on the 16 new proposed members of $\beta$~Pictoris and calculated approximate membership probabilities (Table~\ref{tab.BANYAN}). We used the distances of the systems provided in Table~\ref{tab.results}, our proper motion measurements in Table~\ref{tab.astrometry}, and radial velocities from the literature (for those objects without radial velocity measurements, we assumed the values of their companions).
Although only seven of the 16 pairs showed high-probability memberships to $\beta$ Pictoris (see Table~\ref{tab.BANYAN}), these results should be used with caution because most of the new candidates lack accurate distances or radial velocities.}

None of the new reported wide systems have parallax measurements for both components.
However, the location of the 16 stars (eight primaries and eight secondaries) in the colour-magnitude diagrams suggests that both components are located at similar distances.
Definitive parallactic confirmation of common distance will have to wait until early 2016 with the second {\em Gaia} release.
In the meantime, we can infer the true physical binding of the systems with the computation of the reduced gravitational binding energy.

\begin{table*}[]
\centering
\caption{Membership probabilities for the 16 new $\beta$\,Pictoris candidates using the BANYAN on-line tool.}
\label{tab.BANYAN}
\begin{tabular}{l c l c l} 
   \hline
   \hline
   \noalign{\smallskip}
Simbad                                  &   $\beta$\,Pic        & Highest                       & $V_{\rm r}$                     & Reference             \\
name                                            &  prob. [\%]                   & prob. [\%]                      & [kms$^{-1}$]          & of $V_{\rm r}$        \\
  \noalign{\smallskip}
    \hline
\noalign{\smallskip}
 
2MASS J00193931+1951050         &  83.9 & $\beta$\,Pic                  & --1.7\,$\pm$\,1.0\tablefootmark{P}              & Schlieder et al. 2012a         \\ 
EX Cet                                  &  99.9 & $\beta$\,Pic                  & +41.8\,$\pm$\,0.7                                                       & Soubiran et al. 2013            \\
CD--44 753 B                            &  0.1  & Tuc-Hor (99.4)                        & +12.4\,$\pm$\,1.9                                                       & Kordopatis et al. 2013          \\ 
2MASS J07293670+3554531 &  23.4 & Field (50.8)                  & +10.4\,$\pm$\,0.9\tablefootmark{P}            & Schlieder et al. 2012a  \\      
L 186--67 B                             &  0.0  & Field (100)                   & +40\,$\pm$\,9   \tablefootmark{P}                       & Kordopatis et al. 2013    \\              
2MASS J08274412+1122029 &  16.0 & Field (84.0)                  & +11.2\,$\pm$\,1.7\tablefootmark{P}            & Schlieder et al. 2012a  \\      
HD 82939 A                              &  1.1  & Field (98.9)                  & --0.5\,$\pm$\,0.4                                                       & Gontcharov 2006         \\      
2MASS J12120849+1248050 &  0.0  & Field (99.9)                  & --4.0\,$\pm$\,1.0\tablefootmark{P}            & Schlieder et al. 2012a  \\      
MV Vir C                                &  0.0  & AB Dor (86.0)                 & +0.0\,$\pm$\,0.8\tablefootmark{P}               & Malo et al. 2014a                     \\
2MASS J16170673+7734028 &  0.0  & Field (100)                   & --14.4\,$\pm$\,1.0    \tablefootmark{P}               & Schlieder et al. 2012a  \\  
2MASS J18420483--5554126        &  99.9 & $\beta$\,Pic                  & +1.0\,$\pm$\,0.7\tablefootmark{P}               & Malo et al. 2013                      \\
HD 173167 A                             &  99.9 & $\beta$\,Pic                  & +0.8\,$\pm$\,7.0                                                        & Mo{\'o}r et~al. 2013            \\
HDE 331149 B                            &  95.3 & $\beta$\,Pic                  & --19.2\,$\pm$\,1.1      \tablefootmark{P}               & Schlieder et al. 2012a           \\      
BPS CS 22898--0066              &  0.0  & Field (100)                   & +0.6\,$\pm$\,3.0                                                        & Kordopatis et al. 2013          \\
2MASS J21551738--0046231        &  75.2 & $\beta$\,Pic                  & ...                                                                             & ...                                             \\
2MASS J23301129--0237227        &  91.2 & $\beta$\,Pic                  & --5.3\,$\pm$\,0.2\tablefootmark{P}              & Malo et al. 2014b             \\
\noalign{\smallskip}
\hline
\end{tabular}
\tablefoot{
\tablefoottext{P}{Radial velocity adopted from the primary component.}
}
\end{table*}

\subsection{Projected separations and binding energies}

In Table~\ref{tab.results} we list the angular separations, $\rho$, and position angles, $\theta$, at the 2MASS epoch of observation of the 36 wide pairs in the $\beta$~Pictoris moving group.
Angular separations vary from 8.2\,arcsec for \object{BD--21~1074}A,Ba,Bb to about 1.3\,deg for the triple system AU~Mic--AT~Mic~AB (Luyten 1941; Caballero~2009). %%% Reaching the boundary… II.
Among our new pairs, $\rho$ varies from 10.6\,arcsec to 24.5\,arcmin.

To distinguish between true very wide physical binaries and co-moving pairs of ``single'' stars that belong to the same kinematic group, we computed the reduced gravitational binding energies, $U_g^* = - G \mathcal{M}_1 \mathcal{M}_2 s^{-1}$ (Caballero 2009), of the 36 systems.
With the angular separations and distances, we obtained the projected physical separations, $s$, which vary from merely 100--120\,au for the known pairs WDS~08228--5727 (L~186--67~Aa,Ab,B) and WDS~10596+2527 (\object{HD~95174}~AB) to about 7\,10$^{4}$\,au (0.34\,pc) for the new pair \object{WDS~08290+1125}.
Given the uncertainties in the distance (Table~\ref{tab.sample}), we provide only two significant figures for $s$.

We derived masses $\mathcal{M}_1$ and $\mathcal{M}_2$ from $J$-band absolute magnitudes $M_J$ and the Baraffe et~al. (2015) or Siess et~al. (2000) evolutionary models at 20\,Ma for solar metallicity and the appropriate mass intervals.
When available, we gathered masses of single early type stars and close binaries from the literature (e.g., Strassmeier \& Rice 2000; Neuh{\"a}user et~al. 2002; Caballero 2009; Donati et~al. 2011; Janson et~al. 2012; Elliott et~al. 2015; {Montet et~al. 2015}) or suitable information that allowed us to make a precise derivation (e.g., magnitude differences from adaptive optics or lucky imaging, mass ratios from spectroscopic monitoring -- Chauvin et~al. 2010;  Neuh{\"a}user et~al. 2011; Janson et~al. 2012; Messina et~al. 2014; Bowler et~al. 2015; Elliott et~al. 2015).
Masses range approximately from 2.4\,$M_\odot$ for \object{$\eta$~Tel}~A to well below the substellar boundary for \object{L~186--67}~B with a broad maximum of the distribution at 0.5--1.0\,$M_\odot$.
Derived masses reasonably match those expected from spectral types, when available.
For the sake of completeness, we also list spectral types compiled from a number of sources in Table~\ref{tab.results} (Riaz et~al. 2006; Reid et~al. 2007; Pickles \& Depagne 2010; R.\,Caballero 2012; Janson et~al. 2012; Kraus et~al. 2014; Messina et~al. 2014; Rodr{\'i}guez et~al. 2014; Mason et~al. 2015; I.\,Gallardo \& M.\,G\'omez Garrido, priv. comm.; SIMBAD).

The greatest absolute value of reduced binding energy among the 36 systems in Table~\ref{tab.results}, of $-U_g^*$ = 9800~10$^{33}$\,J, corresponds to the strongly bound pair HD~95174~AB, which is not only the tightest one, but also contains two stars of $\sim$0.8\,$M_\odot$.
On the other hand, there are two very fragile system candidates with binding energies of 0.57--2.7~10$^{33}$\,J, almost one order of magnitude lower than that of the Luyten's system AU~Mic+AT~Mic\,AB, which lies at the boundary between very wide binaries and couples of single stars that are co-moving within the same stellar kinematic group (Caballero 2010; see the title of this series of papers).
As a result, it is likely that the components in the two new fragile system candidates {\em WDS~08290+1125} and {\em WDS~23317--0245}, which includes the flaring star \object{AF~Psc} (Bond 1976; Kraus et~al. 2014; Ramsay \& Doyle 2014), originated in the same parental cloud and were ejected at the same time, in the same direction, and at the same velocity, but they are not physically bound.
The six other new pairs have binding energies between 13 and 1400~10$^{33}$\,J and may survive the eventual disruption by the Galactic gravitational potential for some billion years (Weinberg et~al. 1987; Close et~al. 2007).
In any case, detecting features of youth in the spectra of {\em WDS~08290+1125}~A and~B and the wide M6.0\,V companion candidate to AF~Psc (Reid et~al. 2007) would shed light on their actual membership in the $\beta$~Pictoris moving group.

\subsection{Benchmark objects and probable members in other young moving groups\label{sec.benchmark}}

The 36 wide systems tabulated by us can help to constrain the actual membership of some controversial candidate members in $\beta$~Pictoris:
\begin{itemize}
        \item {WDS~01367--0645}.
Some authors have also classified the primary of the system, \object{EX~Cet} (G5\,V), as a member of the Hercules-Lyra association ($\sim$100--200\,Ma -- Montes et~al. 2001; L\'opez-Santiago et~al. 2006; Shkolnik et~al. 2012; Eisenbeiss et~al. 2013).
        \item {WDS~02305--4342}.
The primary CD--44~753~A is also a member candidate of the Columba association ($\sim$15--50\,Ma -- Torres et~al. 2006, 2008; Elliott et~al. 2014; Malo et~al. 2014a).
        \item {WDS~08228--5727}.
The membership of the primary \object{L~186--67}~Aa,Ab to $\beta$~Pictoris is ambiguous (Malo et~al. 2013, 2014a).
The late-M common proper motion companion, L~186--67\,B, whose physical binding in the system had been confirmed earlier (Bakos et~al. 2002; Bergfors et~al. 2010; Janson et~al. 2012, 2014), would have a mass close to the deuterium-burning mass limit if it were 20\,Ma old.
If membership in $\beta$~Pictoris were confirmed, the triple system would be a benchmark for very low-mass substellar astrophysics.
        \item {WDS~09361+3733}.
While there are no membership studies for the primary, the homonymous secondary \object{HD~82939}~Ba,Bb was listed not only as a $\beta$~Pictoris star by Schlieder et~al. (2012a, 2012b), but also as a young field star by Malo et~al. (2014b).
        \item {\textit{WDS~16172+7734}}.
Schlieder et~al. (2012a) listed the primary \object{TYC~4571--1414--1} as a probable member of both $\beta$~Pictoris and AB~Doradus ($\sim$70\,Ma) moving groups.
        \item {WDS~21214--6655}.
The primary star \object{V390~Pav}~A has also been classified as a member of the Tucana-Horologium association ($\sim$30\,Ma -- Zuckerman et~al. 2001a; Mamajek et~al. 2004; Rojas et~al. 2008).
\end{itemize}

If the six systems above were eventually discarded as true $\beta$~Pictoris ``pairs'', 30 systems would still remain for further investigation in the young moving group, of which six (20\,\%) are reported here for the first time.

Certain systems in Table~\ref{tab.results} are also particularly important in the low-mass domain, because they can be used to test evolutionary models.
Just to cite one example, the secondary of the pairs \textit{WDS~16172+7734} (presented here for the first time) and WDS~21105--2711 (Bergfors et~al. 2010; Malo et~al. 2013, 2014b) lie close to the substellar limit and, therefore, to the lithium depletion boundary.
As a result, a high-resolution spectroscopic analysis of both primaries and secondaries could shed more light on the debated age of $\beta$~Pictoris.

%
%________________________________________________________________
\section{Conclusions\label{sec.conclusions}}

We searched through 35 previous publications and compiled an exhaustive list of 185 members and member candidates in the nearby, young ($\sim$20\,Ma) $\beta$~Pictoris moving group, around which we looked for common proper-motion companions at projected physical separations of up to 10$^5$\,au.
For that, we made extensive use of the Aladin and STILTS virtual observatory tools and numerous public all-sky catalogues (e.g., WDS, PPMXL, 2MASS).

Of the 184 initial common proper-motion companion candidates, 129 were the subject of a precise astrometric follow-up, by which we measured proper motions with typical uncertainties of only 1\,mas/a, and 67 of a multi-band photometric study.
Eventually, we discarded five previously reported pairs and retained 36 reliable pair candidates.
Of them, 18 and 10 are known systems with both components or only one component classified as $\beta$~Pictoris members, respectively, and eight are new pairs in the moving group.
We also report 16 new star and brown dwarf candidates in $\beta$~Pictoris for the first time.
These values represent an increase of 9\,\% in the total number of reported objects in the moving group and of almost 30\,\% in the number of wide proper motion systems.

We investigated the 36 pairs with available public information in detail.
Among them, there are 12 triple and two quadruple systems, which points out to a greater incidence of high-order multiplicity in $\beta$~Pictoris than in the field, possibly ascribed to a member list biased towards close binaries {or an increment of the high-order, multiple fraction for very wide systems.}

We measured angular separations and projected physical separations, compiled or derived masses for components in all systems, and computed reduced gravitational binding energies.
Two of the new pair candidates could be unbound couples of single stars that are co-moving within $\beta$~Pictoris, while at least one of the components in six (new and known) pairs have also been reported to belong to other young moving groups and associations (four in Hercules-Lyra, Columba, AB~Doradus, Tucana-Horologium) or to the field (two).
There are three pairs (one presented here) with masses of secondaries at or below the hydrogen-burning limit, and they can be used as benchmarks for {upcoming} age-dating works in $\beta$~Pictoris.
Our study provides a comprehensive analysis of the wide multiplicity in one of the closest and youngest moving groups known and, therefore, also serves as input to models of moving-group evolution and eventual dissipation by the Galactic gravitational field.

%
%________________________________________________________________
%%% acknowledgment
\begin{acknowledgements}
{We thank the anonymous referee for the report.}
This research made use of the Washington Double Star Catalogue, maintained at the U.S. Naval Observatory, the SIMBAD database and VizieR catalogue access tool, operated at Centre de Donn\'ees astronomiques de Strasbourg, France, and the Spanish Virtual Observatory ({\tt http://svo.cab.inta-csic.es}).
Financial support was provided by the {Universidad Complutense de Madrid}, the Comunidad Aut\'onoma de Madrid, and the Spanish Ministerios de Ciencia e Innovaci\'on and of Econom\'ia y Competitividad under grants
AP2009-0187,                            % (FPU Alonso-Floriano)
AYA2011-24052,                          % proyecto E. Solano
and AYA2011-30147-C03-02, and -03.      % (CARMENES-IAA, UCM, CAB)
\end{acknowledgements}
%
%________________________________________________________________

%
%________________________________________________________________

\appendix
\section{Long tables}
\onecolumn

\begin{longtable}{l c c c l c}
\caption{\label{tab.sample}Investigated $\beta$\,Pictoris members and member candidates.}\\
   \hline
   \hline
\noalign{\smallskip}

 Simbad   			&  $\alpha$ 	& $\delta$	& $d$  		& 	Reference for&  Memb.	   \\
 name				&  (J2000)	& (J2000)	& $\lbrack$pc$\rbrack$ & distance	 &  flag		 \\

\noalign{\smallskip}
    \hline
\noalign{\smallskip}
 \endfirsthead
  \caption{Investigated $\beta$\,Pictoris members and member candidates (cont.).}\\ 
  \hline
  \hline
\noalign{\smallskip}	

 Simbad   			&  $\alpha$ 	& $\delta$	& $d$  		& 	Reference for& Memb. 	   \\
 name				&  (J2000)	& (J2000)	& $\lbrack$pc$\rbrack$ & distance	 & flag 		 \\

\noalign{\smallskip}
  \hline
\noalign{\smallskip}
  \endhead
\noalign{\smallskip}
  \hline
  \endfoot  
  
HD 203                         	& 00:06:50.08	& --23:06:27.2 &  39.39 $\pm$ 0.59	& van Leeuwen 2007		& 1  \\ 
RBS 38                         	& 00:17:23.54	& --66:45:12.5 &  39.0  $\pm$ 2.6	& Riedel et al. 2014  		& 2 \\ 
RX J0019.7+1951		       	& 00:19:43.04	& +19:51:11.7  &  59.4  $\pm$ 7.9	& Schlieder et al. 2012a	& 2   \\%1RXS J001942.4+195126 ,{[SLS2012]} PYC J00197+1951  
FK Psc			       	& 00:23:34.68	& +20:14:28.3  &  53 $\pm$ 4	  	& Malo et al. 2014b		& 2   \\ 
1RXS J002700.0+663025    	& 00:27:02.83	& +66:30:39.0  &  56.9  $\pm$ 8.3	& Schlieder et al. 2012a  	& 3   \\
GJ 2006 A		       	& 00:27:50.23	& --32:33:06.4 &  33.2  $\pm$ 2.8	& Riedel et al. 2014   	  	& 1  \\%
GJ 2006 B                      	& 00:27:50.35	& --32:33:23.9 &  31.5  $\pm$ 2.4	& Riedel et al. 2014  		& 1  \\ 
LP 525--39 AB                  	& 00:32:34.81	& +07:29:27.1  &  41.1  $\pm$ 4.4	& Schlieder et al. 2012b	& 2   \\
\end{longtable}

%
%________________________________________________________________________________________________________________________________________________

\newpage
\begin{longtable}{l l l c c l}
\label{tab.WDS}\\
\caption{Unresolved or unidentified systems.}\\
   \hline
   \hline
\noalign{\smallskip}

Simbad   	&	 Discovery name		&  Status	& $\rho$		& $\theta$	& WDS				\\	  
name		&	 or reference    	&  						& [arcsec]				& [deg]	& identifier		\\	 

\noalign{\smallskip}
    \hline
\noalign{\smallskip}
 \endfirsthead
  \caption{Unresolved or unidentified systems (cont.).}\\ 
  \hline
  \hline
\noalign{\smallskip}	

Simbad   	&	 Discovery name		&  Status	& $\rho$		& $\theta$	& WDS				\\	  
name		&	 or reference    	&  						& [arcsec]				& [deg]	& identifier		\\	
\noalign{\smallskip}
  \hline
\noalign{\smallskip}
  \endhead
  \hline
  \endfoot  
\noalign{\smallskip}
LP 525--39 AB		& MCT 1				& Physical	& 0.7		& 334	& 00326+0729	\\ 	
\noalign{\smallskip}	
RBS 153 AB			& BRG 3				& Physical	& 0.4		& 168	& 01072--1936	\\ 	
\noalign{\smallskip}
LP 467--16 AB		& BEU 2				& Physical	& 0.3		& 241	& 01114+1526	\\ %LP 467--16 AB	
\noalign{\smallskip}
2E 327 AB			& BRG 4				& Physical	& 1.4		& 28	& 01135--3821	\\ 	
\noalign{\smallskip}
LP 648--20 BC		& BRG 5				& Visual	& 5.4		& 184	& 01376--0645	\\ %(Not Physical)	
LP 648--20 BD		& BWL 7				& Visual	& 6.7		& 23	& 			\\% 
\noalign{\smallskip}
BD+17 232AB		& COU 254				& Physical	& 1.6		& 24	& 01377+1836 	\\
\noalign{\smallskip}
RBS 253 AB			& BRG 7				& Physical	& 2.8		& 292	& 01538--1500	\\
\hline

\end{longtable}

%
%________________________________________________________________________________________________________________________________________________

\scriptsize

\begin{landscape}

\begin{longtable}{l c c c c c c c c l l}
\caption{Astrometry measurements.\label{tab.astrometry}}\\
   \hline
   \hline
\noalign{\smallskip}
Simbad   			&  $\alpha$ 	& $\delta$	& $\mu_{\alpha} \cos \delta$	 	& $\mu_{\delta}$	 	& $\mu_{\alpha} \cos \delta$	& $\mu_{\delta}$ 	& $\Delta$t		& Number of & Origin	 of adopted& Astrometric	\\
name				&  (J2000)	& (J2000)	& [mas\,a$^{-1}$]			&[mas\,a$^{-1}$]		& [mas\,a$^{-1}$]		&[mas\,a$^{-1}$]	& [a]			& detections	& proper motions	& status	\\
				&  		&		& \multicolumn{2}{c}{PPMXL}						& \multicolumn{2}{c}{Adopted}										\\

\noalign{\smallskip}
    \hline
\noalign{\smallskip}
 \endfirsthead
  \caption{Astrometry measurements (cont.).}\\ 
  \hline
  \hline
\noalign{\smallskip}	
Simbad   			&  $\alpha$ 	& $\delta$	& $\mu_{\alpha} \cos \delta$	 	& $\mu_{\delta}$	 	& $\mu_{\alpha} \cos \delta$	& $\mu_{\delta}$ 	& $\Delta$t		& Number of	& Origin of	adopted& Astrometric	\\
name				&  (J2000)	& (J2000)	& [mas\,a$^{-1}$]			&[mas\,a$^{-1}$]		& [mas\,a$^{-1}$]		&[mas\,a$^{-1}$]	& [a]			& detections	& proper motions & status	\\
				&  		&		& \multicolumn{2}{c}{PPMXL}						& \multicolumn{2}{c}{Adopted}										\\

\noalign{\smallskip}
  \hline
\noalign{\smallskip}
  \endhead
\noalign{\smallskip}
  \hline
  \endfoot

RX J0019.7+1951		      			& 00:19:43.04   & +19:51:11.7   & +69.8 $\pm$ 5.1	& --49.8 $\pm$ 5.1   & +72.6 $\pm$ 0.2 		& --43.9 $\pm$ 0.3	& 56.586  & 6 	&{\em This work}		&		\\
2MASS J00193931+1951050    			& 00:19:39.31   & +19:51:05.1   & +70.3 $\pm$ 4.1	& --48.9 $\pm$ 4.1   & +74.7  $\pm$ 0.4 	& --40.9 $\pm$ 0.1	& 56.586  & 6 	&{\em This work}		& Physical (in WDS)		\\%1RXS J054223.7--275803
\noalign{\smallskip}
GJ 2006 A		      			& 00:27:50.23   & --32:33:06.0  & +96.3 $\pm$ 22.0	& --101.9 $\pm$ 22.0 & +108.5 $\pm$ 0.4		& --47.8 $\pm$ 0.8	& 27.188  & 5 	&{\em This work}		&    		\\
GJ 2006 B		      			& 00:27:50.35   & --32:33:23.8  & +117.1 $\pm$ 20.7	& --45.7 $\pm$ 20.7  & +101.8 $\pm$ 0.3		& --42.2 $\pm$ 1.0	& 27.188  & 4 	&{\em This work}		& Physical (in WDS)		\\
\noalign{\smallskip}
LP 648--20 (EX Cet B)	      	      			& 01:36:55.17   & --06:47:37.9  & +180.3 $\pm$ 4.1  	& --105.4 $\pm$ 4.1  & +166.8 $\pm$ 0.3		& --99.9 $\pm$ 0.3	& 11.682  & 5   &{\em This work}	 	& 	    	\\ 
EX Cet		      			& 01:37:35.47   & --06:45:37.5  & +171.2 $\pm$ 0.8  	& --97.7 $\pm$ 0.6   & +171.3 $\pm$ 0.7  	& --98.5 $\pm$ 0.5	& ...	  & ... & HIP2				& Phot. follow-up   	\\ %     
\noalign{\smallskip}
BD+17 232 AB		      			& 01:37:39.39   & +18:35:32.7   & +69.9 $\pm$ 1.5	& --47.5 $\pm$ 1.5   & +69.9 $\pm$ 1.5  	& --47.5 $\pm$ 1.5  	& ...  	  & ...	& TYC   			&    		\\ 	  
2MASS J01372935+1805559       			& 01:37:29.35   & +18:05:55.9   & +73.4 $\pm$ 4.1	& --45.1 $\pm$ 4.1   & +73.4 $\pm$ 0.1    	& --46.2 $\pm$ 0.1  	& 55.603  & 5	&{\em This work}		& Phot. follow-up	\\
\end{longtable}

\end{landscape}
%
%________________________________________________________________________________________________________________________________________________

\scriptsize

\begin{landscape}

\begin{longtable}{l c c c c c c c l}
%\label{tab.A3}\\%
\caption{\label{tab.photometry}Photometry measurements.}\\%
   \hline
   \hline
\noalign{\smallskip}
Simbad   			&  $\alpha$ 	& $\delta$	& {\em B}	& {\em V}	& {\em r'}	& {\em K$_{s}$} & d		& Photometric	\\%
name				&  (J2000)	& (J2000)	& [mag]		& [mag]		& [mag]		&  [mag]	& [pc]		& status	\\%

\noalign{\smallskip}
    \hline
\noalign{\smallskip}
 \endfirsthead
  \caption{Photometry measurements (cont.).}\\% 
  \hline
  \hline
\noalign{\smallskip}	
Simbad   			&  $\alpha$ 	& $\delta$	& {\em B}	& {\em V}	& {\em r'}	& {\em K$_{s}$} & d		& Photometric	\\%
name				&  (J2000)	& (J2000)	& [mag]		& [mag]		& [mag]		&  [mag]	& [pc]		& status	\\%

\noalign{\smallskip}
  \hline
\noalign{\smallskip}
  \endhead
\noalign{\smallskip}
  \hline
  \endfoot

 RX J0019.7+1951 				   & 00:19:43.04 & +19:51:11.7 & 17.075 $\pm$  0.700 &  ...                & 14.694 $\pm$  0.064 &  9.864 $\pm$  0.018 & 59.40 $\pm$  7.90 	& 			     \\% 
 2MASS J00193931+1951050			   & 00:19:39.31 & +19:51:05.1 & 17.079 $\pm$  0.010 & 15.540 $\pm$  0.010 & 14.913 $\pm$  0.040 & 10.078 $\pm$  0.018 & 		   	& Physical (A in $\beta$\,Pic)	     \\% 
\noalign{\smallskip}
 GJ 2006 A					   & 00:27:50.23 & --32:33:06.0 & 14.333 $\pm$  0.050 & 12.869 $\pm$  0.100 & 12.312 $\pm$  0.100 &  8.012 $\pm$  0.033 & 33.2 $\pm$ 2.80 	& 			     \\% 
 GJ 2006 B					   & 00:27:50.35 & --32:33:23.8 & 14.659 $\pm$  0.080 & 13.165 $\pm$  0.030 & 12.605              &  8.116 $\pm$  0.027 & 		 	& Physical (A and B in $\beta$\,Pic)			     \\% 
\noalign{\smallskip}
 LP 648--20 (Ex Cet B)				   & 01:36:55.17 & --06:47:37.9 & 15.725 $\pm$  0.050 & 13.996 $\pm$  0.110 & 13.385 $\pm$  0.080 &  8.862 $\pm$  0.021 & 23.95 $\pm$  0.42 	& 			     \\% 
 EX Cet 					   & 01:37:35.47 & --06:45:37.5 &  8.661 $\pm$  0.990 &  7.750 $\pm$  0.990 &  7.635 $\pm$  0.001 &  5.752 $\pm$  0.018 & 			& Physical candidate (B in $\beta$\,Pic)          	     \\% 
\noalign{\smallskip}
 BD+17 232AB					   & 01:37:39.39 & +18:35:32.7 & 11.062 $\pm$  0.060 &  9.890 $\pm$  0.020 &  9.600 $\pm$  0.010 &  6.716 $\pm$  0.018 & 52.59             	& 			     \\% 
 2MASS J01372935+1805559 			   & 01:37:29.35 & +18:05:55.9 & 16.826 $\pm$  0.200 & 15.454 $\pm$  0.340 & 14.621 $\pm$  0.030 & 10.734 $\pm$  0.022 & 			& Visual          	     \\
\end{longtable}

\end{landscape}

%
%________________________________________________________________________________________________________________________________________________

\onecolumn
\scriptsize
\begin{landscape}

\begin{longtable}{l l l c c c c l c c c c c l}
\caption{\label{tab.results} Common proper-motion companion candidates.}\\
   \hline
   \hline
\noalign{\smallskip}

WDS	& Discovery	& Simbad   			&  $\alpha$ 	& $\delta$	& $d$		       & $M_{J}$	       & Sp.	       & $\rho$        & $\theta$      & s     & $\mathcal{M}$ & $\textit{--U}^{*}_{g}$        & Comments\\
identifier	& name$^{a}$    & name				&  (J2000)	& (J2000)	& [pc]		       & [mag]  	       & type	       & [arcsec]      & [deg]         & [10$^{3}$\,au] &[$M_{\odot}$]  &   [10$^{33}$\,J]      &    \\

\noalign{\smallskip}
    \hline
\noalign{\smallskip}
 \endfirsthead
  \caption{Common proper-motion companion candidates (cont.).}\\ 
  \hline
  \hline
\noalign{\smallskip}	

WDS	& Discovery	& Simbad   			&  $\alpha$ 	& $\delta$	& $d$	     		& $M_{J}$			& Sp.		& $\rho$ 	& $\theta$	& s 	& $\mathcal{M}$ & $\textit{--U}^{*}_{g}$	& Comments\\
identifier	& name$^{a}$    & name				&  (J2000)	& (J2000)	& [pc]       		& [mag]			& type		& [arcsec]	& [deg]		& [10$^{3}$\,au]	&[$M_{\odot}$]  &   [10$^{33}$\,J] 	&    \\
  \hline
\noalign{\smallskip}
  \endhead
\noalign{\smallskip}
  \hline
  \endfoot
\multicolumn{12}{c}{Our common proper-motion candidates}\\
\noalign{\smallskip}
\hline
\noalign{\smallskip}
01376--0645	& 		& EX Cet 			& 01:37:35.47	& --06:45:37.5	& 23.95 $\pm$ 0.42	& 4.32 $\pm$ 0.04	& G5\,V		&		&	& 	& 0.92	&  &					\\ % 
	   	& CAB 3		& LP 648--20 (EX Cet B)			& 01:36:55.17	& --06:47:37.9	& 			& 7.81 $\pm$ 0.04	& M3.5\,V	& 612.1 	& 259	& 15		& 0.12	& 13		& In AF11, 1 new star, Her-Lyr? 	\\ %
\noalign{\smallskip}
\it{02017+0117} &		& {[SLS2012]} PYC J02017+0117N& 02:01:46.78	& +01:17:16.2 	& 63.7 $\pm$ 9.0	& 5.1  $\pm$ 0.3 	& ...		&		&	&	& 0.73						\\
		&New		& {[SLS2012]} PYC J02017+0117S	& 02:01:46.93 	& +01:17:06.0 	&			& 5.1  $\pm$ 0.3	& ...		& 10.48		& 167 	& 0.67   & 0.71		& 1400 & ...		\\
\noalign{\smallskip}
\it{05200+0613}	& 	& TYC 112--917--1		& 05 20 00.29 	& +06 13 03.6	& 68.5	& 5.15 $\pm$ 0.03	& K3		&		&	& 	& 	0.70 	&				\\ %
		& New		&  2E 1249 AB			& 05 20 31.83	& +06 16 11.5	& 		& 5.07 $\pm$ 0.02	& K3		& 506.2  	& 248 	& 35	& 	0.66+0.15	& 28 	& ...			\\ %	
\noalign{\smallskip}
\it{08290+1125}& 		& [SLS2012] PYC J08290+1125 	& 08:29:04.12	& +11:25:05.4	&  58.8 $\pm$ 8.5	& 5.7 $\pm$ 0.3	& ...		&		&	& 	& 0.50						\\
\end{longtable}

\end{landscape}

%
%________________________________________________________________________________________________________________________________________________

%
%________________________________________________________________

%
%________________________________________________________________

\end{document}